\documentclass[fleqn,usenatbib]{mnras}
\usepackage{newtxtext,newtxmath}

\usepackage[T1]{fontenc}
\usepackage{ae,aecompl}

\usepackage{graphicx}	

\usepackage{amssymb}	
\usepackage{pifont}     

\newcommand{\kms}{ {\rm km~s\textsuperscript{-1}}}

\newcommand{\oversim}[2]{\protect{\mbox{\lower0.5ex\vbox{%
   \baselineskip=0pt\lineskip=0.2ex
   \ialign{$\mathsurround=0pt #1\hfil##\hfil$\crcr#2\crcr\sim\crcr}}}}}
\catcode`\"=\active\let"=\"

\def\3{{\ss} }

\def\c12{{1\over 2}}

\def\plusplus{\raise 0.3ex\hbox{${\scriptstyle ++}$}{}}

\def\and{{{\rm M}31}}

\newcommand{\seq}{\,{=}\,}

\defcitealias{Petersen..commensurabilities..2021}{Paper I}
\defcitealias{Petersen..torque..2019}{Paper II}

\def\aj{AJ}
\def\apj{ApJ}
\def\apjl{ApJL}
\def\mnras{MNRAS}

\def\aap{A\&A}

\def\kms{ {\rm km~s\textsuperscript{-1}}}

\usepackage{soul}

\begin{document}

\title[Bar dynamical length]
 {Measuring the dynamical length of galactic bars}

\author[Petersen, Weinberg, \& Katz]
{Michael~S.~Petersen,$^{1}$\thanks{michael.petersen@roe.ac.uk} Martin~D.~Weinberg,$^2$ Neal~Katz$^2$\\ $^1$Institute for Astronomy, University of Edinburgh, Royal Observatory, Blackford Hill, Edinburgh EH9 3HJ, UK \\ $^2$Department of Astronomy, University of Massachusetts at Amherst, 710 N. Pleasant St., Amherst, MA 01003}

\maketitle
\begin{abstract}
We define a physically-motivated measure for galactic bar length, called the {\it dynamical length}.
The dynamical length of the bar corresponds to the radial extent of the orbits that are the backbone supporting the bar feature.
We propose a direct observational technique using integral field unit spectroscopy to measure it.
Identifying these orbits and using the dynamical length is a more faithful tracer of the secular evolution and influence of the bar.
We demonstrate the success of the metric for recovering the maximal bar-parenting orbit in a range of simulations, and to show its promise we perform its measurement on a real galaxy.
We also study the difference between traditionally used ellipse fit approaches to determine bar length and the dynamical length proposed here in a wide range of bar-forming \(N\)-body simulations of a stellar disc and dark matter halo.
We find that ellipse fitting may severely overestimate measurements of the bar length by a factor of 1.5-2.5 relative to the extent of the orbits that are trapped and actually comprise the bar.
This bias leads to overestimates of both bar mass and the ratio of corotation radius to bar length, i.e. the bar speed, affecting inferences about the evolution of bars in the real universe.
\end{abstract}

\begin{keywords} galaxies: Galaxy: halo---galaxies: haloes---galaxies: kinematics and dynamics---galaxies: evolution---galaxies: structure \end{keywords}

\section{Introduction} \label{sec:introduction}

A bar feature in a disc galaxy is made up of stars on particular orbits.
Barred galaxies with stable bars must include the parenting $x_1$
family of orbits \citep{contopoulos80, binney08}.
The $x_1$ orbits can come in a variety of flavours, but the salient
features are the same: elongated orbits with coherent apocentres.
It is these coherent apocentres that give the strong `bar' appearance.

In observations, one typically measures stellar bar structures in
galaxies by fits using simple ellipses \cite[see e.g. ][for a large
sample]{kruk18}.
Using isophotes, one places a series of increasing semi-major axis
ellipses on the image of a barred galaxy.
One uses the `best-fit' ellipse, chosen based on some criteria such
as the ellipse with the maximum ellipticity, to inform the a
fundamental property of a bar: its length.
Unfortunately, simulators have also seized on ellipse fits as a means
to measure and compare galaxies\footnote{Some studies, both observational
and theoretical, use surface density Fourier measures to constrain the
size of the bar; these measurements suffer from the same sorts of biases
and ambiguities as ellipse fits.
See, e.g. \citet{Frankel..TNG..2022} for a comparison between simulations
and observations using Fourier measurements.}.
For galaxy evolution studies, simply computing the ellipse-measured bar
statistics and looking for relationships with other galaxy properties may
confound, or at worst, mislead inferences about the nature of galaxy evolution.
Further, many different ellipse-determination schemes exist in the literature.

Bars are composed of orbits that are trapped in their own potential and
ellipse fitting can include material that is not trapped and therefore not
part of the bar.
In fact, as we show in later sections, such material can overwhelm ellipse
fitting methods.
Hence, it is not clear what information the ellipse measures contain about
the {\it dynamics} of the system.
A variety of works have sought connections between observational
characteristics of ellipse-measured bars and galaxy evolution.
The general picture is that (1) bars are fast, that is, their length is some
appreciable fraction of the corotation radius of a galaxy
\citep{perez12, GarmaOehmichen..MWanalogues..2022}, (2) bars are redder
than their host discs \citep{kruk18}, and (3) bars are long-lived, both
from studies of the Milky Way \citep{bovy19} and observations of
high-redshift bars \citep{hodge19}.

The recent proliferation of studies using integral field units (IFUs) to
image barred galaxies has ushered in a new era, where the kinematics of
a barred galaxy can be studied.
Various works have used the Tremaine-Weinberg method
\citep{Tremaine.Weinberg.patternspeed.1984} and IFU data to measure
the pattern speed of bars
\citep[see the recent data sample for MaNGA galaxies in][]{Guo..MaNGA..2019}.
The typical finding is that bars rotate `fast', signified by the ratio of the
corotation radius to the length of the bar being significantly larger than unity
\citep{debattista98}.
The theoretical relevance of `fast' vs `slow' or more generally the ratio
\({\cal R}\equiv R_{\rm CR}/R_{\rm bar}\) is that one requires
\({\cal R}\ge1\) for
orbits to reinforce a bar potential and, therefore, be trapped. Empirically,
bars in simulations tend to form with \({\cal R}\approx 1\) and change owing to
torques thereafter.
However, resolved stellar velocity fields from IFUs hold more information about
the orbital content of a galaxy.
The $x_1$ orbit family has a unique velocity signature, in that the maximum
radial turning points (apsides) coincide with the bar major axis.

Given that the `fast versus slow' bar dichotomy is based on measuring the length
of the bar, the length of the bar has particular importance for studies of bar
evolution. In this paper, we present a diagnostic that can measure the maximal
extent of the $x_1$ orbits by looking for a unique velocity signature.
We calibrate the technique against an evolving barred-galaxy simulation that has
already been extensively studied in
\citet[][hereafter Paper I]{Petersen..commensurabilities..2021} and
\citet[][hereafter Paper II]{Petersen..torque..2019}.

This paper is organised as follows.
In Section~\ref{sec:methods}, we introduce the dynamical length of the bar and
briefly describe the simulation, along with relevant findings for the simulation
from previous work \citepalias{Petersen..commensurabilities..2021}.
In Section~\ref{sec:barmeasurement}, we compare global measurements of the
simulation: a direct comparison of ellipse fitting and dynamical measurements from
orbital analysis.
This section also presents a kinematic technique with which we determine the true
length of orbits that form the backbone of the bar.
We apply the measurement methods to an observed galaxy in
Section~\ref{sec:examples}.
We discuss the findings in the context of galaxy evolution studies in
Section~\ref{sec:discussion} and conclude in Section~\ref{subsec:conclusions}.
Appendix~\ref{sec:signaloptimisation} explores other variations of our approach
to determine the dynamical bar length, Appendix~\ref{sec:ellipsefitdemo} explores
a wider range of ellipse fitting techniques, and Appendix~\ref{sec:samplegalaxies}
applies our method to a wide range of bar simulations to show its robustness.

\section{The dynamical length of a bar} \label{sec:methods}

To validate our later application to IFU data, we first apply the presented
techniques to a simulation of a disc galaxy that forms a bar.
One may apply the techniques to any simulation of a disc galaxy that forms a
bar, and the results can be validated as long as one can extract a potential.
For this work, we use a previously-analysed simulation from
\citetalias{Petersen..commensurabilities..2021}: the `cusp' simulation.
As we evolved using the basis function expansion $N$-body code {\sc exp}
\citep{weinberg99,Petersen..EXP..2022}, extracting the potential for further
analysis is straightforward through a representation by basis functions.

The simulation we consider features an exponential stellar disc embedded
in a live NFW dark matter halo.
When describing the disc throughout this work, we will work in unitless
`scale length' units, using the initial scale length of the exponential stellar
disc, $R_d$, to facilitate the translation between simulations and observations.
For a Milky Way-like galaxy, a scale length is approximately 3 kpc.

\subsection{The theoretical basis for the dynamical length}

The maximum extent of trapped orbits that comprise the backbone of the bar
\citep[the $x_1$ family;][]{contopoulos80,binney08}, sets the
{\it dynamical length} of the bar.
If one were to axisymmetrise the $x_1$ orbits in a given barred galaxy,
the bar would disappear, whereas anything outside of the $x_1$ family
can be shuffled without a significant reduction in the self-gravity of the bar.
Furthermore, outside of the bar dynamical length, the orbits may {\it gain}
angular momentum (from the bar), resulting in a phase-lagged spiral structure.
This is in contrast to the classic secular evolution picture where bar
orbits {\it lose} angular momentum, resulting in a bar pattern speed decrease (slowing).
The material outside of the dynamical length responds to the $x_1$ orbit
family, appearing as an elliptical distortion of untrapped stars lingering
near the bar.
This material would not remain `bar-shaped' in the absence of the trapped
$x_1$ orbits.
Such orbits can greatly change the inferred bar lengths using ellipse fits.

The dynamical length of a bar is a more informative metric for the evolution
of the bar, as it more closely approximates the self-gravity in the bar feature.
Using the dynamical length, and hence the truly trapped orbits, to estimate the
mass of the bar will also result in a more accurate estimate for the bar mass,
something that we will return to in future work.

The dynamical length of the bar can provide constraints on both the pattern speed
and the mass distribution of the inner galaxy.
This amounts to finding the maximal $x_1$ orbit, which may be possible to directly
determine through kinematics.
The dynamical length can also help determine the location of resonances that may
feed the bar, i.e. add mass and length to the bar through the trapping of new
orbits.

\subsection{Identification of bar-parenting orbits in self-consistent potentials}
\label{subsubsec:trappedanalysis}

In \citetalias{Petersen..commensurabilities..2021}, we analysed bar membership in
this simulation through the clustering of the radial turning points, or apsides,
for a given orbit.
Orbits `trapped' by the bar's gravity will librate about the position angle of
the bar's major axis, and will appear as (nearly) closed orbits in the frame of
the bar.
In this work, we primarily considered orbits that contribute to the structure of
the bar: the $x_1$ family associated with the inner Lindblad resonance (ILR;
$2\Omega_\phi-\Omega_r \seq 2\Omega_p$, where $\Omega_p$ is the pattern frequency
of the bar).
These orbits comprise the `backbone' of the bar and are eccentric orbits
elongated along the bar axis.

\setcitestyle{notesep={ }}
Briefly described \citepalias[see][for details]
{Petersen..commensurabilities..2021}, our method isolates the turning points in
an orbit by looking for local maxima in radius.
\setcitestyle{notesep={, }}
Using a rolling average of 20 apsides\footnote{We determined the rolling average
of 20 empirically to be a sweet spot in a trade-off between time resolution and
signal-to-noise.} in Cartesian coordinates that we transform to a frame co-
rotating with the bar position angle, \(\theta_{\rm bar}\), we compute the
position angle for the centre of two \(k\)--means--derived clusters relative to
the bar, taking the maximum of the two values.
The desire to align with the two ends of the bar motivates our choice of two
clusters. In addition to the cluster position angles, we compute the variance in
the position angle relative to the cluster centre over the 20 apsides,
\(\sigma_{\theta_{\rm bar,20}}\).
These two quantities alone allow for a successful classification of orbits into
the \(x_1\) family if we limit the average apse position to be
$\langle\theta_{\rm bar}\rangle \le \pi/8$ and the variance to be
$\sigma_{\theta_{\rm bar,20}}\le \pi/16$.
From an empirical examination of the orbits, we estimate a contamination
rate in the $x_1$ family of approximately 1 per cent.
This uncertainty does not change any of the results we present in this work.

\begin{figure*} \centering \includegraphics[width=6.5in]{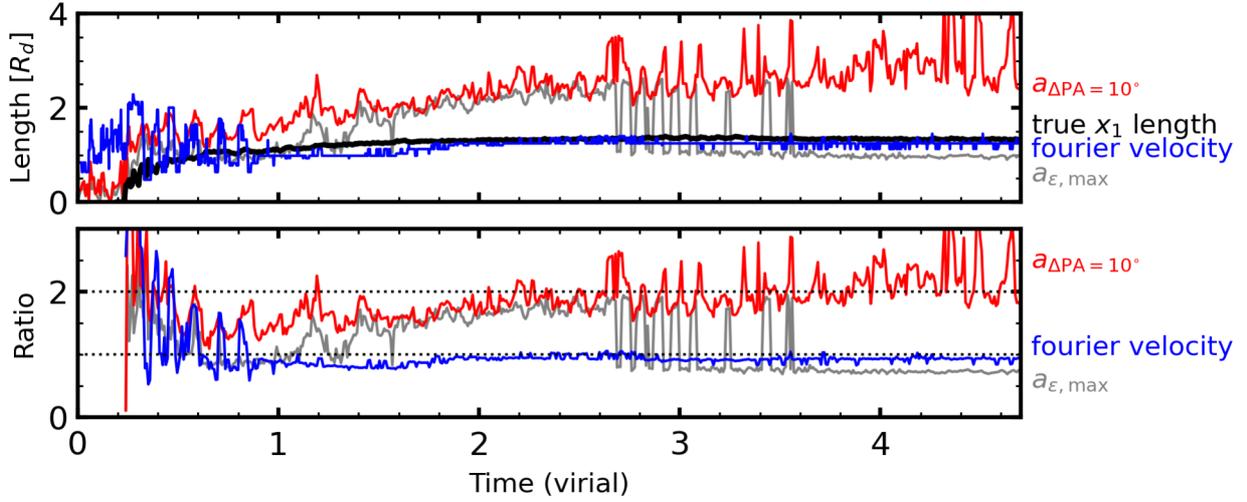}
\caption{\label{fig:ellipse_fit}
Upper panel: The length of the bar in disc scale lengths, measured using four
different techniques: maximal $x_1$ extent, two different ellipse fits, and our
$x_1$ velocity diagnostic, versus time. The $x_1$-derived length is the black curve.
The simulation ellipse-fit-derived length from maximum ellipticity is shown in
grey and the position angle length is in red. The $x_1$ velocity diagnostic is
shown in blue. Lower panel: The ratio of the ellipse-fit-derived length to the
$x_1$--derived length versus simulation time for the measurement techniques in
the upper panel (same colour scheme). The $x_1$ velocity diagnostic is the least
biased measure of the true $x_1$ length.} \end{figure*}

We plot the turning radius of the longest $x_1$, identified using the above
method, as the black curve in the upper panel of Figure~\ref{fig:ellipse_fit}.
In \citetalias{Petersen..commensurabilities..2021} we identified three stages of
bar evolution in this simulation: the shortlived assembly phase when the bar
first forms, the growth phase when the bar grows in the length and mass, and the
steady-state phase when the bar does not evolve much at all.
The length of the $x_1$ orbits grows rapidly through the first time unit
(assembly phase), slowly through the second (growth phase), and essentially
not at all after that (steady-state phase).
All together, the dynamical length appears to show little evolution over the
bulk of the simulation.
However, we emphasise that this model is missing several elements of realism
for comparison to real galaxies, and should, therefore, be treated instead as a
testing ground for theoretical insights, such as the identification of the $x_1$
population.
In the next section, we present measurements of the bar in all three phases (at
$T=0.6, 2.0, 4.0$), to explore when one expects our method to work well. We investigate additional simulations in a phase-agnostic manner in
Appendix~\ref{sec:samplegalaxies}.

With the importance of $x_1$ orbits in mind, as well as the observationally-unfortunate necessity of a potential model to conclusively identify them, we endeavour to find an {\it indicator} that can track the extent of
the maximal $x_1$ orbit in a observationally-friendly and computationally inexpensive manner.

\section{Observational Bar Measurement Techniques} \label{sec:barmeasurement}

In this section, we compare two methods to parameterise the size of the bar:
measurement of the maximum radius of the trapped $x_1$ orbits that support the bar potential
and traditional ellipse fitting to the isophotes of surface density.

\begin{figure*} \centering \includegraphics[width=6.5in]
{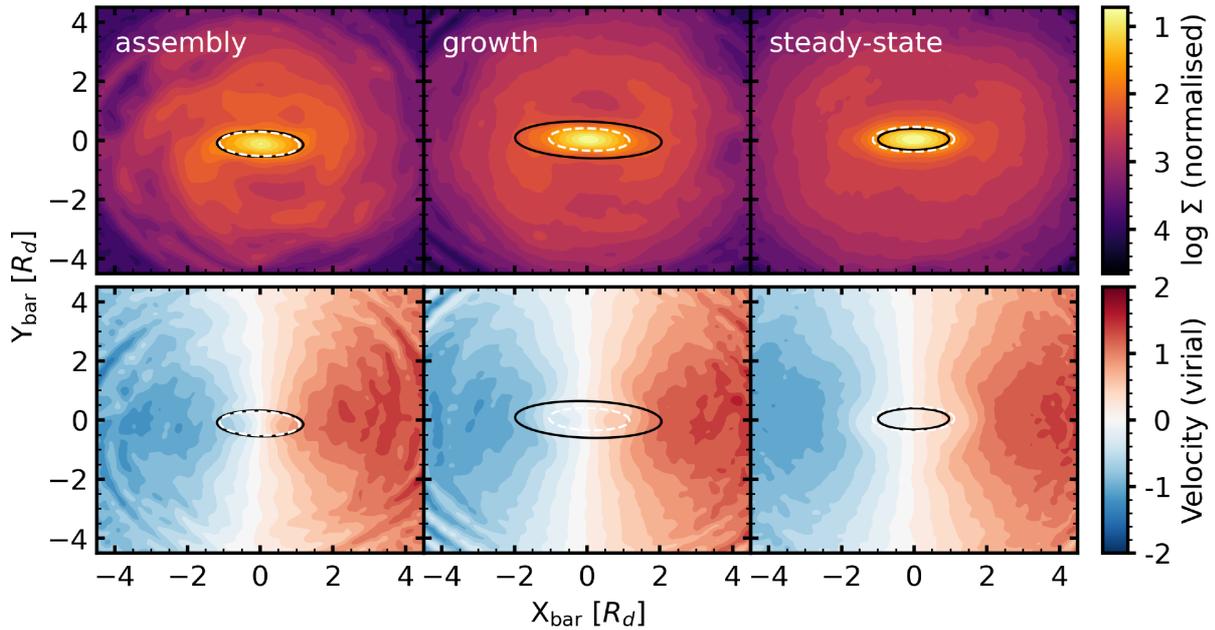}
\caption{\label{fig:cusp_velocity}
Upper panels: Log surface density, in
normalised units, for the three evolutionary phases in the simulation:
assembly, growth, and steady-state. Lower panels: The velocity field in the
direction tangential to the bar, for the three phases in the upper panels.
The white dashed ellipses show the maximum
extent of the trapped $x_1$ orbits, which coincides with the dimple in the
velocity field in the growth and steady-state phases, as calculated from $v_{4\perp}/v_{2\perp}$ for the velocity field tangential to the bar (see
text). The black dashed ellipses show the maximum ellipticity ellipse for the
bar from the surface density plot alone.} \end{figure*}

\begin{figure} \centering \includegraphics[width=3.2in]
{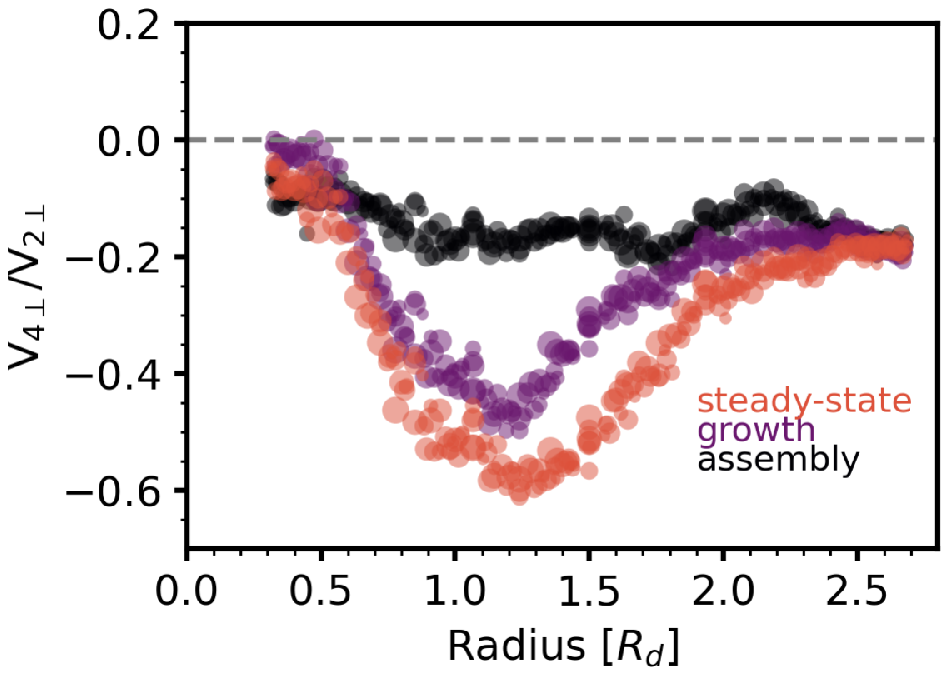}
\caption{\label{fig:fourierdynamicallength} Demonstration of the $x_1$ velocity method for finding the dynamical length of a bar. Each coloured set
of circles, corresponding to repeated samplings using different binning, is the Fourier indicator computed for a different phase of the bar evolution.
The outer envelope of the circles gives a measure of the uncertainty. The size of
the circle corresponds to the root variance of the individual observation.} \end{figure}

\subsection{Bar Measurement via Kinematic Signatures}\label{subsec:kinematics}

\citetalias{Petersen..commensurabilities..2021} demonstrated that $x_1$ orbits
dominate the population of the bar, particularly at later stages of bar
evolution. In turn, this suggests that such orbits may also dominate integrated
kinematic measurements. Using our theoretical understanding of $x_1$ orbits,
we can make predictions for the velocity signatures of these orbits. Because the
tangential velocity of the bar orbits reach a minimum at the end of the
bar,
the
velocity
should show a marginally lower amplitude at the end of the bar than that
expected for ordinary disc orbits.
Figure~\ref{fig:cusp_velocity} shows the surface density
(upper panels) and velocity fields (lower panels) for the three model
potentials studied in this paper, confirming by eye that the velocity field
features `dimples' as a result of the bar. The dimples are an $m=4$ signal that
arises from the libration signature of the $x_1$ orbits. The ability to
distinguish a trapped $x_1$ orbit from an untrapped orbit will be largest at
the extrema of the libration oscillation.  In other words, the dimple is an arc of libration
turnarounds. In the case of the perfect parent $x_1$ orbit, this feature would
be a point, but for librating orbits the dimples smear into an arc with
four-fold symmetry.

Our developed technique harnesses this feature as a diagnostic of the
maximal extent of $x_1$ orbits. Given a continuous velocity field tangential
to the bar axis (such as from an integral field unit), the signal from trapped
$x_{1}$ bar orbits will be low compared to disc orbits and will, therefore, be
largest where the discrepancy between the bar orbits and the disc orbits is
largest,
i.e. at the `four corners' of the bar. This suggests that a kinematic metric
using four-fold $m\seq4$ symmetry will reveal the largest differences between
the bar and disc velocities. The difference between the velocities of bar and disc orbits will be negative at the corners of the bar and so we expect the $m\seq4$
velocity moment tangential to the bar,
\begin{equation}
    v_{4\perp}=\frac{1}{N}\sum_{i=0}^N v_{\perp,i}\cos\left(4\phi_i\right),
    \label{eq:m4velocity}
\end{equation}
to be appreciably negative. The quantity $v_{\perp,i}$ is the velocity
perpendicular to the bar for an individual particle or pixel; in our test cases
where the bar is aligned with the $x$ axis, the perpendicular velocity is simply
$v_y$. As the bar slows, we expect the $v_{4\perp}$ quantity to become even more extreme since the velocity between the untrapped disc orbits
and the bar pattern speed becomes more discrepant
\citepalias{Petersen..commensurabilities..2021}. In
Equation~(\ref{eq:m4velocity}), we compute the sum for a set of particles;
here we choose particles in annuli to obtain $v_{4\perp}(R)$ as a function of
cylindrical radius.

To test the significance of the signal relative to that of the bar, we compare
the $m\seq2$ and $m\seq4$ velocity moments, $v_{4\perp}$ to
\begin{equation}
    v_{2\perp}=\frac{1}{N}\sum_{i=0}^N v_{\perp,i}\cos\left(2\phi_i\right).
\end{equation}
This suggests a procedure to observationally determine the maximal extent
of $x_1$ orbits, which we call the {\it $x_1$ velocity method}, as follows:
\begin{enumerate}
\item Compute the {\it magnitude} of the velocity perpendicular to the bar,
$v_\perp$, for each particle $i$.
\item Compute the $m\seq2$ and $m\seq4$ $x_1$ velocity components as a function
of radius, $v_{2\perp}(R_j)$ and $v_{4\perp}(R_j)$, partitioning the particles
into bins delineated by annular radii $\{R_j\}$. One must take care not to
reduce the signal-to-noise ratio by choosing annuli that are too narrow
relative to the spatial resolution\footnote{In practice, we estimate the
uncertainty on the values of $v_{2\perp}(R_j)$ and $v_{4\perp}(R_j)$ via a
jackknife procedure, taking $\sqrt{N_j}$ samples in each bin, where $N_j$ is
the number of particles in the bin, and compute the root variance for each set
of Fourier coefficients.}. We suggest a minimum annular radius of
$\delta r \seq 10\delta x$, where $\delta x$ is the pixel scale.
\item Locate the radius for the
${\rm min}\left\{v_{4\perp}(R_j)/v_{2\perp}(R_j)\right\}_j$.
As long as ${\rm min}\left\{v_{4\perp}(r_j)/v_{2\perp}(r_j)\right\}_j$ is
significant\footnote{We do not currently have an quantitative measure of the
significance, but in practice we have found the minima are unambiguous. We
discuss strategies for estimating the uncertainty below.}, this method is
reliable and the minimum approximately equals the maximal extent of the $x_1$
orbits.
\end{enumerate}

To use this metric, one requires high spatial and velocity resolution, coupled
with a modest inclination. As a general guideline, this requires a spatial
resolution of $\delta x\seq0.05R_d$, where $R_d$ is the disc scale length, for a
galaxy at an inclination of 45$^\circ$, and requires a velocity resolution of
$0.05v_{\rm max}$, where $v_{\rm max}$ is the maximum circular velocity, to
determine the $v_{4\perp}/v_{2\perp}$ metric.  For a MW-like galaxy, this
translates to an $\approx$10$\kms$ velocity resolution and a 150 pc spatial
resolution.

In Figure~\ref{fig:fourierdynamicallength}, we show this diagnostic for the model
potentials using the above method. As expected, all three models show negative
$v_{4\perp}/v_{2\perp}$ values, driven by the effect at the corners of the bar.
As the diagnostic requires binning to compute $v_{4\perp}/v_{2\perp}$, we sample
each model with a range of bin widths, increasing the number of bins. For each
measurement in a different bin, we measure the root variance via a jackknife
procedure, repeatedly randomly sampling the particles in the bin\footnote{This
will not be strictly possible in real observations, but we perform the procedure
here to test the robustness of the binning.}. The bins are then shown as circles
whose radii correspond to the root variance of the individual observation. The
outer envelope of the circles gives an approximate measure of the uncertainty in
determining the underlying curve.

The signal is very strong during the growth and steady-state phases, with only a
weakly discernible minima signature near the end of the bar in the assembly
phase. One expects a low signal in the assembly phase since the kinematic feature
results from trapped, evolved orbits that develop at the start of the growth
phase.  Even in the case of marginally significant detections, the
$v_{4\perp}/v_{2\perp}$ method results in a more accurate bar length
determination compared to methods using ellipse fits.
While we restrict our measurements to the $m=2$ and $m=4$ components of the
velocity field, we have also tested the inclusion of higher order harmonics to
improve the quality of detections, as described in
Appendix~\ref{sec:signaloptimisation}.

In the upper panel of Figure~\ref{fig:ellipse_fit}, we plot the length of the bar
derived using the $x_1$ velocity method as a blue curve.
In the lower panel of Figure~\ref{fig:ellipse_fit}, we compute the ratio of this
estimated length to the actual maximal $x_1$ length and it accurately finds the
true maximal $x_1$ orbit length.
Hence, this technique is a simple way to
probe for the dominant barred galaxy orbit and determine a
dynamically-relevant length for a bar, and also a more accurate measure of the
trapped fraction in galaxies.
The technique does require a continuous velocity field, but these are becoming readily available with IFUs.
In Appendix~\ref{sec:samplegalaxies}, we apply our $x_1$ velocity method to a wide sample of barred galaxy simulations to confirm that our method remains robust.

\subsection{Bar Measurement via Ellipse Fits}\label{subsubsec:ellipsefits}

To place our results in context, we compare with ellipse fits to bar isophotes.
Many studies fit ellipses to bar isophotes \citep[e.g.][]{munozmateos13,
laurikainen14, kim15, erwin16, kruk18}\footnote{While some papers have recently
moved to Fourier-based decompositions of the bar \citep[e.g.][]
{Guo..MaNGA..2019,Frankel..TNG..2022}, such measurements of bars are subject to
the same biases and algorithmic-consistency challenges as ellipse fits, so the
general spirit of this section still holds.}.
The various ellipse measurements have known discrepancies.
In an example, \citet{athanassoula02} demonstrate that different ellipse methods
applied to the same galaxy can lead to variations of up to 35 per cent in the
measured bar length.

Following on this work, \citet{munozmateos13} measured bar lengths in the S$^4$G
sample \citep{sheth08} using four different ellipse metrics derived from either
the ellipticity profile or position angle of the best-fit ellipse at a given
radius, reproduced here for clarity:
(1) $a_{\varepsilon, {\rm max}}$, the radius where the ellipticity of the bar is
maximum,
(2) $a_{\varepsilon, {\rm min}}$, the first local minimum after the maximum,
(3) $a_{\Delta\varepsilon=0.1}$, the radius where the ellipticity drops by 0.1
compared to the maximum ellipticity, and
(4) $a_{\Delta {\rm PA}=10^{\circ}}$, the radius where the position
angle differs by $>$10$^\circ$ from the position angle at the maximum
ellipticity.

The range of options for classifying the length of the bar highlights the
challenge of algorithmically finding a bar length based on ellipse
fits\footnote{Many studies have made use of visually-determined bar lengths,
including Galaxy Zoo \citep{hoyle11} and S$^4$G \citep{sheth08}. While visually
measuring a bar length is a quick process for individual galaxies, the method
does not necessarily trace an isodensity surface and often offers no errors on
individual measurements. \citet{hoyle11} found that individual observers report
approximately a 6 per cent deviation relative to the mean of all observers who
classify a bar length. Connecting to ellipse-fit methods, \citep{herreraendoqui15}
demonstrated that visual fits to bar lengths are roughly comparable to
$a_{\epsilon, {\rm max}}$.}. It is generally claimed in the literature that
$a_{\epsilon, {\rm max}}$ may be thought of as a minimum bar length \citep[e.g.][]
{athanassoula02, martinezvalpuesta06,munozmateos13}. Therefore, some studies have
used the maximum ellipticity to identify bars. For instance, in a recent paper
studying bars in JWST early release fields \citep{Guo..CEERSbars..2022}, the
maximum ellipticity value defines the bar length.  However, no study that we have
identified to-date has performed orbit decompositions to find the true dynamical
backbone of the bar. And as we shall see below, even the assumed lower limits for
bar length are overestimates of the maximal $x_1$ length.

To perform a comparison between measurement techniques, we fit ellipses to our
simulated galaxy. We use {\tt elliptical}, a purpose-built ellipse fit Python
package that implements many procedures common to ellipse fits of barred
galaxies. In brief, {\tt elliptical} accepts an image, then uses {\tt scikit-
image.find\verb|_|contours} \citep{skimage}, a marching-squares algorithm, on the
face-on image of the galaxy to determine the position of isophotal contours given
a surface density. Then {\tt elliptical} computes the best fit ellipses from the
matrix representation of conic sections\footnote{Alternately, one can perform a
least-squares fit to the contour level \citep[e.g.][]{athanassoula90}, which
provides an additional level of flexibility in fits, but we do not find a
substantial difference in our ellipse fit measurements if we choose this
technique versus the matrix representation.} for a range of isophotal values in
the vicinity of the bar. Given the set of fit ellipses, we can select different
criteria for the bar length. Examples are given in
Appendix~\ref{sec:ellipsefitdemo}. In practice, we select the maximum ellipticity
as the bar length, for reasons that will become clear below. As a second
comparison, we select an alternate bar length as the smallest ellipse where the
position angle has changed by 10$^\circ$ from the position angle at maximum
ellipticity. In the upper panel of Figure~\ref{fig:ellipse_fit}, we plot these as
the grey curve (maximum ellipticity length) and the red curve (position angle-
derived length). In the lower panel of Figure~\ref{fig:ellipse_fit}, we compute
the ratio of the ellipse-derived measures to the maximal $x_1$ length.

We find that selecting different criteria such as the maximum ellipticity or a
threshold in position angle variation does not qualitatively change our results;
all ellipse measures from \citep{munozmateos13} return values within
approximately 30 per cent, in agreement with \citet{athanassoula02}. For a more
thorough introduction to measuring bar lengths, we refer the reader to
\citet{erwin05} for an observationally motivated viewpoint, and
\citet{athanassoula02} for a theoretically motivated viewpoint.

\subsection{Why do the techniques return different results?}
\label{subsec:dressing}

If bars are made of elongated, coherent orbits, and orbits make the elongated,
coherent bar structures that may be traced by ellipse fits, why do the two
methods give different results? In this section, we describe orbits that bias the ellipse fits of bars and
demonstrate how ellipse fitting methods may overestimate the length of the bar.

Figure~\ref{fig:cusp_velocity} summarises our velocity moment method as compared
to standard ellipse fits. The upper panels of Figure~\ref{fig:cusp_velocity} show
the surface density during the three identified phases of bar evolution, and the
lower panels show the $y$-component of the velocity (perpendicular to the bar)
for each of the three phases.

In each phase, we plot the standard technique (ellipticity drop) best-fit
ellipse in dashed black. The ellipse that corresponds to the minimum of
$v_{4\perp}/v_{2\perp}$ is shown in dashed white. As shown in
Figure~\ref{fig:fourierdynamicallength}, while we compute a minima in the
$v_{4\perp}/v_{2\perp}$ value for the assembly phase, the signal is fairly
ambiguous, despite still giving a matching value to the maximal $x_1$
length\footnote{With the omniscience provided by simulations and the true
calculation of the maximal $x_1$ orbits, we see that the $x_1$ track is not
yet fully formed (the apsis precession that traps orbits into the bar is an
ongoing
process), and thus the technique will not, by definition, be informative.}.
However, during the growth and steady-state phases, strong signals are
observed in $v_{4\perp}/v_{2\perp}$, which cleanly show the maximal extent of the
$x_1$ orbit family. The discrepancy between ellipse-fit lengths and
the actual maximal $x_1$ orbit radius can be significant.

Assuming a constant mass-to-light ratio, we apply the standard ellipse-fitting
analysis to our simulations. We compute the face-on surface density at a
resolution of $0.025a$, which for a MW--like galaxy corresponds to 75
pc\footnote{In practice, the ellipse fits are weakly
dependent on the resolution, introducing approximately a 10 per cent error.}.
We measure the length of the bar using a standard method: determine best-fit
ellipses at many different surface densities and assign the bar length to the
semi-major axis value, $a$, where the ellipticity drops below a certain
threshold or has a discontinuity. Here, we choose the semi-major axis where
the ellipticity $e\equiv1-\frac{b}{a}$, and $b$ is the semi-minor axis value,
first drops below 0.5. In practice, this is also the same location as the
discontinuity in $e$ that corresponds to the transition between bar--dominated
contours and disc--dominated contours. We also have a dynamically-informed
metric for the length of the bar: the maximum extent of the $x_1$ family
\citepalias{Petersen..commensurabilities..2021}. Figure~\ref{fig:ellipse_fit}
shows a comparison of the maximum $x_1$ extent versus ellipse--fit derived bar
lengths. In the upper panel, we show the bar lengths measured using both
techniques. The two ellipse fit lengths in the simulation (grey and red curves) grow steadily at $T<1$, except for a later shift to a new set of ellipses from
the maximum ellipticity measurement ($T>3$). The growth roughly mirrors the trapped fraction growth, but when compared to the maximal extent of the $x_1$
orbits (black), we see that this is an extreme overestimate for the length of
the trapped component. The periodicity in the ellipse measurements results
from the outer disc $m\seq2$ disturbances, e.g. spiral arms, coincidentally
aligning with the
bar. At early times when the bar is forming, this can result in variations of
nearly a factor of two. Even at later times in the simulation, the variation
in ellipse--fit length is 25 per cent over short ($\delta T\seq0.1$)
timescales owing to the $m\seq2$ alignment.

\begin{figure*} \centering \includegraphics[width=6.5in]
{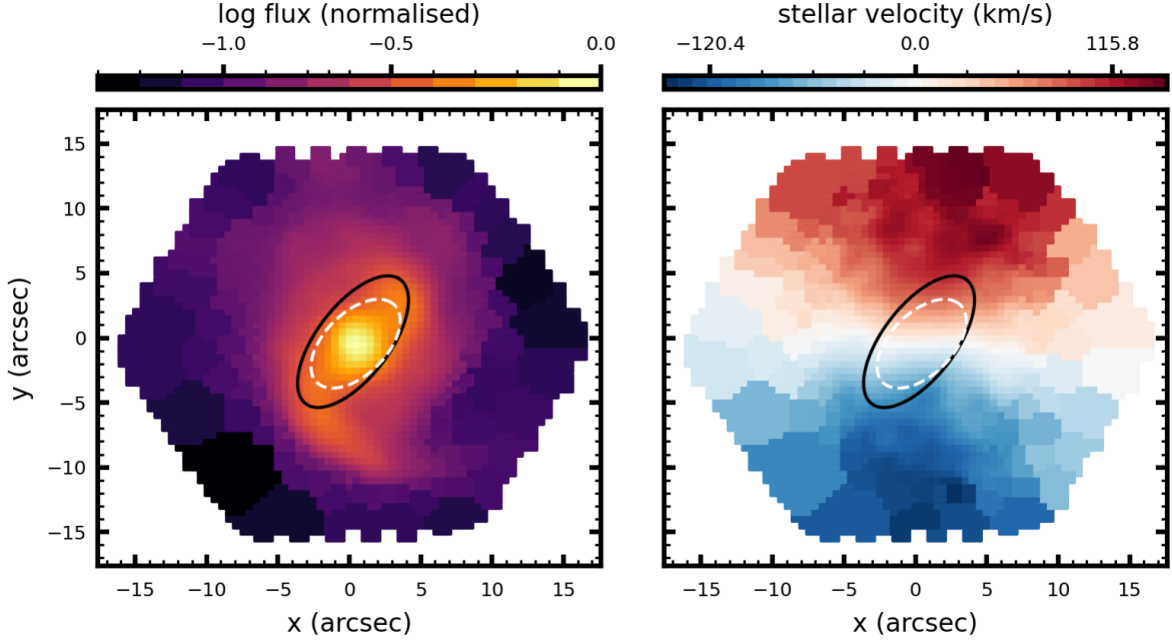}
\caption{\label{fig:manga844412703}
Left panel: log surface density, normalised to the central value, for the MaNGA
galaxy 8444-12703. North is up, east is right (note that many display images are
flipped with respect to the $y$-axis). Right panel: stellar velocity, in km/s. We
show two ellipses in both panels: the black solid ellipse is the maximum-
ellipticity ellipse drawn from a sample of ellipses fit to the isophotes. The
white dashed ellipse is drawn from the same sample of ellipses fit to isophotes,
but has been chosen as the ellipse with the semi-major axis that most closely
matches the measured dynamical length of the bar.} \end{figure*}

The lower panel of Figure~\ref{fig:ellipse_fit} summarises the overall results of
our comparison. We plot the ratio of the ellipse-fit length to the maximal $x_1$
length for the simulation. The ellipse-fit length is a large overestimate for the
length of the $x_1$ orbits at all times in the simulation, typically by a factor
of 1.5, except during the assembly phase when the overestimate is a factor of
two. Thus, the simulation reveals the ambiguity in ellipse-fit determinations of bars.
Ellipse fits may be taken as a measure of the total mass distribution of the
galaxy, not as the mass directly associated with the bar itself.

The gravity from the bar causes the untrapped disc orbits to linger along the bar
position angle. This confuses metrics that are based on the
shape alone, such as ellipse fits and Fourier-derived strengths.
Typically, the length of
the bar will be overestimated by 50 per cent. One needs an accurate length of
the true bar orbits to observationally determine the pattern speed \citep[e.g.]
[]{Guo..MaNGA..2019,GarmaOehmichen..MWanalogues..2022}. The dimensionless
parameter $\mathcal{R}\equiv R_{\rm CR}/R_{\rm bar}$ denotes the `slowness' of the pattern speed. As discussed
in \citet{binney08}, this parameter can be hard to measure in real galaxies
for two reasons: (1) the bar does not have a sharp end, and (2) corotation
does not have a clear definition for strong non-axisymmetric disturbances,
e.g. a strong bar. However, several studies have attempted to measure either
the pattern speed or the dimensionless parameter to characterise the bar.
Thus, an overestimate of the true dynamical length of the bar overestimates
the pattern speed, sometimes significantly.

In Appendix~\ref{sec:samplegalaxies}, we apply the ellipse fitting technique to a
wide sample of barred galaxy simulations and confirm that these conclusions still
hold. Other studies have found similar results, but for somewhat different
reasons: \citet{Hilmi..etal..2020} points out that ellipse fits may encompass
both bar and spiral activity, confusing dynamical information. Additionally the
spiral arms may create real modulations in the bar feature, particularly the
pattern speed. We do not find any obvious evidence for pattern speed modulation,
and we can conclusively say that the length of the $x_1$ orbits does not
fluctuate on spiral-arm timescales when spirals are present in our model
(primarily during the growth phase).

In conclusion: {\it if one must pick an ellipse fit measurement for the bar
length, the maximum ellipticity is likely to be the best tracer of the $x_1$
orbits.}

\section{Example applications} \label{sec:examples}

\subsection{Utility for IFU surveys}

As a proof of concept example, we apply our measurement technique to a Milky-Way-
like galaxy  measured in the MaNGA survey \citep{Bundy..MaNGA..2015}. We select
galaxy 8444-12703, a part of a Milky-Way-analogue subsample
\citep{GarmaOehmichen..MWanalogues..2022}. We perform the analysis purely on the
MaNGA data cube. Using the log flux, we draw ellipses on the image.  This bar is particularly interesting, as both a Milky Way analogue, but also for the
unphysically-low measurement of the dimensionless bar pattern speed ratio
$\mathcal{R}=0.9\pm0.2$
\citep{GarmaOehmichen..MWanalogues..2022}. This low $\mathcal{R}$ measurement
owes to finding a large bar length of $5.7\pm0.9$ arcseconds. This reported bar
length corresponds to the maximum ellipticity radius, which is similar to the
$5.8$ arcsecond value returned by {\tt elliptical} when computed on the MaNGA log
flux image. We show the ellipse of constant surface density, measured by {\tt
elliptical}, as a black ellipse in both panels of Figure~\ref{fig:manga844412703}.

Using the deprojection values from \citep{GarmaOehmichen..MWanalogues..2022}, and
using {\tt elliptical} to perform the deprojection, we rectify the velocities to
be face on, with the bar aligned with the $x$-axis to find the velocity field
perpendicular to the bar. Using this velocity field, we compute the velocity
Fourier moments as a function of radius\footnote{As a measure of uncertainty, we
sample the bin spacing several times and plot the resultant points scaling
according to the root variance of individual measurements (as in Figure~\ref{fig:fourierdynamicallength}). The envelope of these
measurements may be used to estimate the uncertainty. In practice, we find that a
conservative estimate of the uncertainty is 15 per cent.}. However, as this fit
has been performed in the deprojected space, we must re-project the bar length
onto the sky image. We do this using the inverse transformation matrix. Given the
reprojected bar length, we identify the ellipse measured by {\tt elliptical}
which has the same semi-major axis length. This is shown as the white-dashed
ellipse in both panels of Figure~\ref{fig:manga844412703}.

We find that our $x_1$ velocity indicator suggests a maximal $x_1$ length of
$4.1\pm0.3$ arcseconds, in contrast to the $5.8$ arcseconds of the maximum
ellipse. If we take the corotation radius at face value from
\citet{GarmaOehmichen..MWanalogues..2022}, this change in bar length immediately
raises the ratio $\mathcal{R}$ to a more physical value, 1.3. Hence, our proof of
concept example demonstrates that when IFU data is present, the stellar velocity
field may be used to additionally help constrain the dynamical length of the bar
and its relationship with corotation.

An analysis of a larger sample of galaxies is beyond the scope of this work; in
fact, this measurement may not be possible for a number of MaNGA galaxies owing
to unavoidable limitations in their orientation (relating to uncertainties in
projection), as well as instrumental effects such as the size of the IFU fibre
field, the coarseness of the spatial sampling, correlated information between
pixels, and signal-to-noise. Future work will have to perform a careful analysis
of these considerations when applying our $x_1$ velocity method to a wider range
of galaxies, and automation may not be trivial. Here, we simply provide a proof-of-concept and evidence that our findings from a model galaxy appear to be
applicable to Nature. We expect that the largest variation in our results
compared to observational findings is likely the uncertainty in the inclination
and the position angle of the line of nodes. Cursory tests suggest that the
introduction of the inclination and position angle of the line of nodes can cause
a 20 per cent variation in the determinations of the maximal $x_1$ orbit length.

\begin{table}
\centering
  \begin{tabular}{|l|c|c|}
Parameter & Literature Value & This paper\\
    \hline
    \hline
$\log\left(\frac{M}{{\rm M}_\odot}\right)$ & 10.9 & -\\
Bar Length (max. $\epsilon$) & 5.7 (arcsec)& 5.8 (arcsec) \\
Bar Length (dynamical) &  - & 4.1$\pm$0.3 (arcsec) \\
Deprojected Bar Length & 5.0$\pm$0.7 (kpc)& 3.6$\pm$0.5 (kpc)\\
Disc PA & $4^\circ.0\pm0.6$ & -\\
inclination & $37^\circ.2\pm1.2$ & -\\
Bar angle & $31^{\circ}.7\pm9.5$ & -\\
$\mathcal{R}$ & $0.9\pm0.2$ & $1.3\pm0.3$\\
\end{tabular}
  \caption{Parameters for MaNGA galaxy 8444-12703. The literature values come from \citet{GarmaOehmichen..MWanalogues..2022}, which measures the mean of the bar length to be $R_\varepsilon$, the maximum ellipticity. The uncertainty is drawn from a lognormal distribution using $R_{\rm PA}$ (the radius at which the position angle changes by $5^\circ$) as the upper limit. 
  Deprojected bar length follows the procedure in \citet{Gadotti..bardeprojection..2007}. }
  \label{tab:manga}
\end{table}

Looking ahead, higher velocity and spatial resolution, potentially feasible
using current instruments, makes a modest difference. Comparison with maximal
$x_1$ extents \citepalias[computed in][as shown in
Figure~\ref{fig:ellipse_fit}]{Petersen..commensurabilities..2021} shows that
the minima of $v_{4\perp}/v_{2\perp}$ is within 10 per cent of the maximal
$x_1$ orbit radius, making this technique a powerful descriminator of the
dynamically-relevant maximal $x_1$ orbit.

\begin{figure*} \centering \includegraphics[width=6.5in]{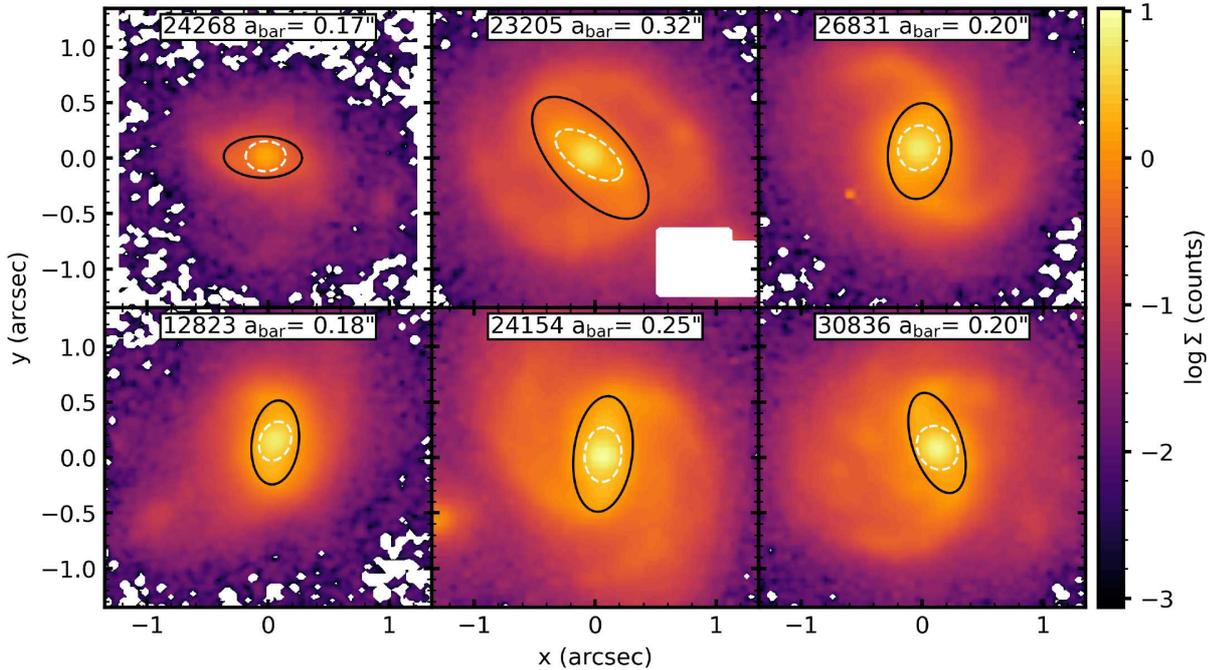}
\caption{\label{fig:ceersbars} Six barred galaxies identified in
\citet{Guo..CEERSbars..2022}, with black ellipses indicating the maximum
ellipticity measure. White dashed ellipses are the best-fit ellipse that has half
the semi-major axis length of the black ellipse, in order to guide the eye as to
possible bar length overestimations (see Section~\ref{subsec:highredshift}).
Regions with negative flux are shown in white. We have masked a
background galaxy in 23205 in order to show the galaxy with clearer contrast.}
\end{figure*}

\subsection{Insights for high-redshift bars}
\label{subsec:highredshift}

Given that the maximum ellipticity appears to be the most robust of the still
biased ellipse based methods, as a tracer of the $x_1$ population, we suggest
that analysis of barred-galaxy images would do better to trace the maximum ellipticity contour as a
measure of the bar length. As a demonstration of how this measure may change
published results, we measure the maximum ellipticity radius for six galaxies
recently identified in JWST imaging as the highest-redshift bar candidates
\citep{Guo..CEERSbars..2022} using {\tt elliptical}. We show the maximum-
ellipticity ellipses overlaid on thumbnail images in
Figure~\ref{fig:ceersbars}.

If we assume a similar potential down-scaling ratio from the maximum
ellipticity fits and $x_1$ lengths, we can attempt to re-calibrate the bar
length, with large uncertainties. Interpreting the range of overestimates from
ellipse fits, we can estimate that some bar lengths may be overestimated by a
factor of 2. To guide the eye on the factor of two reduction, we also add a
constant surface density tracing ellipse with half the semimajor axis of the
maximum ellipticity ellipse as a dashed white ellipse. Several of these
contours are quite circular; for some bars this may owe to the pixel size of
JWST, but for others this may indicate that the bar feature is weak or
potentially non-existent.

The galaxies in the CEERS dataset are prime candidates to observe using the integral field spectroscopy capabilities of JWST. Such analysis, seeking to
identify $x_1$ orbits in particular, are a necessary step in analysing the
evolution of bars over cosmic time, possibly enabling statements about the
evolutionary phase and determining whether these bars are created via
instabilities in early-Universe discs, or whether the bars are more likely to
be directly caused by interactions.

\section{Discussion} \label{sec:discussion}

\subsection{Comparison with fixed-potential analyses}

The key quantity for the purposes of the analysis in this paper is the
determination of the maximal $x_1$ orbit turnaround radius.
From fixed potential analyses, one can compute the theoretical maximum of the
$x_1$ family \citepalias[e.g.]
[]{Petersen..commensurabilities..2021}. However, in self-consistent simulations,
orbits will not necessarily reach the theoretical maximum, but will instead only
extend to some fraction of it, as orbits near the theoretical maximum are
susceptible to small perturbations that untrap them. This clouds the
straightforward interpretation of fixed-potential analyses. The fixed-potential
integration finds possible
$x_1$ orbits out to 2$R_d$, while the maximal $x_1$ found in the self-
consistent simulation at the same time is 1.3$R_d$, a significant
difference\footnote{Knowing where the maximum $x_1$ orbit is located in $r_{\rm
apocentre}-v_{\rm apocentre}$ space, we can efficiently check our trapping
analysis to ensure that we have not missed any elongated $x_1$ orbits. We find
that the $k$-means analysis efficiently recovers the most extended $x_1$
orbits in the simulation, which are {\it not} the same extension as the
theoretical maximum $x_1$ orbit.}. In light of this, methods that parameterise
the length of the bar as the maximal theoretical $x_1$ extent will overestimate
the actual bar length \citep{martinezvalpuesta06}.

Therefore, while a fixed-potential analysis has utility for characterising the
possible orbital content, it is a more informative diagnostic of model bar
evolution to probe the true extent of the $x_1$ orbits. This is particularly
important for comparison to observations, since one can only observe the maximum
length of populated $x_1$ orbits.

\subsection{Use in cosmological simulations}
\label{subsec:cosmo}

The identification and measurement of bars in cosmological simulations is a
similar, but not identical problem to the isolated model we study here. In the
absence of the ability to resolve individual orbits, our $x_1$ velocity method
may be straightforwardly applied to cosmological simulations to both
detect and parameterise bars. Given particle data, one may easily compute the
Fourier velocity moments. We have not investigated the utility of our Fourier
velocity method on particle-based data, but we expect that the method could be
readily calibrated.

One such velocity-based bar measurement was attempted in
\citet{Reddish..newhorizon..2022}, but without a robust sample of bars, the
exercise was more useful as a validation that the method can be used in tandem
with surface density Fourier methods to detect the presence or absence of bars.
In general, we endorse the practice of using velocity-based measurements to probe
the presence of bars in simulations and observations, as examining the kinematics
can reveal bar-supporting orbits, while also preventing false positives from
ellipse fits to surface density.

Finding $x_1$ orbits in cosmological simulations will be limited by temporal
resolution. That is, the ability to resolve time-dependent orbits in cosmological
simulations is typically a simple issue of the time between snapshots. Since the
size of cosmological simulation outputs is large, the time spacing is usually
large. However, a cosmological simulation invested in resolving the dynamics of
bars could readily implement an on-the-fly $x_1$ orbit finder using the apocentre
clustering techniques presented in this paper and in
\citetalias{Petersen..commensurabilities..2021}. However, there is a concern that
noise in the potential calculation in cosmological simulations, i.e. the forces,
may cause unphysical orbit diffusion and hence further tests will be needed to
see if our $x_1$ velocity approach leads to valid bar measurements in these
circumstances.

\subsection{Comparison of observations and simulations}

Given the prevalence of bars in galaxy surveys and the recent ability of
cosmological simulations to form bars \citep{Frankel..TNG..2022}, developing a
common measurement system with which to interpret observations is paramount.
However, one should not degrade the information in simulations to match the
observations; rather one should work to find tracers of dynamically-relevant
features in observations. In this work, we present a consistent dynamical measure
for bars in simulations: the maximum extent of $x_1$ orbits. We demonstrate that
the maximum extent of $x_1$ orbits may be recovered via the velocity field, and
in particular using simple Fourier expansion measurements of the velocity field.

Simple ellipse fits (or surface density Fourier expansions) will be biased by
disc orbits that are deformed (but not trapped) in response to the $x_1$-parented
potential. These deformed orbits are not accurate tracers of bar evolution, and
indeed appear to play little role in the evolution of model galaxies
\citepalias{Petersen..torque..2019} apart from feeding the growing bar
\citepalias{Petersen..commensurabilities..2021}.

In a volume- and mass-limited study of local galaxies \citet{erwin19} found a
bimodal correlation between bar semimajor axis and stellar mass: at lower masses
bar size is independent of stellar mass while at higher masses bar size strongly
depends on the stellar mass. The dependence of stellar mass on galaxy size cannot explain the bar sizes at higher masses. More likely than not, these observations
are confounded by their bar size definitions that depend on integrated light
rather than orbit properties.

Similarly, over cosmic time, bar lengths strongly correlate with galaxy mass, but seem to show little to no evolution with redshift over $0<z<0.84$
\citep{Kim.etal.2021}. Again, this may indicate that ellipse bar measurements are
obscuring the true dynamical evolution.

In concert with a mass model, the maximum $x_1$ extent can also be used to
constrain the pattern speed of a barred galaxy. The idea of using the maximal
extent of $x_1$ orbits to characterise the Milky Way bar was recently studied by
\citet{Lucey..mwbar..2023}. However, one must take care when comparing mass
models with observations, as the theoretical maximum $x_1$ orbit is unlikely to
be populated.

\section{Conclusion} \label{subsec:conclusions}

We introduced the concept of the {\it dynamical length} of a galactic bar: the maximum
extent of the $x_1$ orbit turnaround radius, the same orbits that parent the bar.
The dynamical length provides a physically-motivated diagnostic that may be
robustly compared across simulations and with observations, provided that the
$x_1$ orbits can be identified. With this motivation, we present an observational
metric based on Fourier expansions of the galaxy velocity field to identify the
maximum extent of $x_1$ orbits in simulations and observations, which we call the {\it $x_1$ velocity method}. IFU stellar
velocity data enables locating orbits trapped in the bar feature using a simple
Fourier-based velocity diagnostic, $v_{4\perp}/v_{2\perp}$. The bar-length bias
caused by orbits deformed by the $x_1$ potential but not actually trapped, can be
mitigated through the inclusion of velocity data. The measurements above work
equally well at all phases of barred galaxy evolution for a wide range of
simulated bars.

We determined the dynamical length in a model galaxy and calibrated the
observational metric against it. From this comparison, we determined that
observational techniques currently used on both real galaxies and simulated
galaxies are not measuring true bar lengths. We perform ellipse fits on our
simulations, finding that typical ellipse-fit techniques systematically
overestimate the maximum radial extent of the trapped bar orbits. We describe why
ellipse based techniques overestimate the length of the bar. Such methods do not
accurately represent the radial extent nor mass of orbits trapped in the bar, and
are subject to unpredictable biases. As a proof of concept we successfully apply
our new $x_1$ velocity method to an observed galaxy to measure its dynamical bar length and as predicted ellipse fitting methods overestimate the true dynamical
bar length.

There is scope for significant future work based on the technique presented here.
In particular, while we applied our method to one example galaxy, large samples
of barred galaxies have IFU data. Such work with samples of galaxies will
undoubtedly have growing pains, and we expect to refine the method with further
input for observational considerations. One additional line of study that may
prove fruitful would be the combination of the indicators presented in this paper
and the Tremaine-Weinberg method for pattern speed measurement
\citep{Tremaine.Weinberg.patternspeed.1984}, which also uses the velocity field.
Also desirable is a bin-free approach to finding the velocity-derived length.
Currently we are using a `brute force' approach to test different bin
configurations, but a continuous function estimator would be much cleaner.

In the context of isolated simulated galaxies, we have not solved the problem of
how to diagnose a bar in the `assembly' phase; neither the velocity-field method
nor $x_1$ indicators work particularly well. However, the assembly phase is
shortlived so our $x_1$ velocity method should work well on most galaxies. With a
focused study of bar formation in simulations, more robust $x_1$ diagnostics
during the assembly phase may be developed (see
Appendix~\ref{sec:signaloptimisation} for a discussion). We can, however,
confidently assess that standard ellipse-fit measurements, including the maximum
ellipticity, are overestimating the length of the bar during the assembly phase.

The observational techniques presented in this paper provide a promising avenue
with which to study dynamical quantities for galactic bars. In turn,
understanding bars may shed light on galaxy evolution as a whole as observations
to increasingly higher redshift have shown that structure formation is
commonplace in the early universe \citep{hodge19,Guo..CEERSbars..2022}.

\section*{Data Availability}
The {\tt elliptical} code, which generated many of the measurements, is freely
available on GitHub. The MaNGA and JWST data were both drawn from public data
sets. Access to the simulation data will be provided upon reasonable request to
the corresponding author.

\section*{Acknowledgments}
MSP acknowledges support from a UKRI Stephen Hawking Fellowship. This work used
cuillin, the Intitute for Astronomy's computing cluster
(http://cuillin.roe.ac.uk), partially funded by the STFC. We thank Eric Tittley
for continued smooth operations. This project made use of {\it numpy}
\citep{numpy} and {\it matplotlib} \citep{matplotlib}. We used {\tt MARVIN}
\citep{Cherinka..MARVIN..2019} to access the MaNGA data cubes.

\bibliography{DynamicalLength}

\appendix

\section{Optimising the signal detection}
\label{sec:signaloptimisation}

As the $x_1$-driven velocity signal is fundamentally a deviation from the
quadrupole structure, seeking to find the four `corners' of the bar in
velocity space, one might naturally wonder whether a data-driven solution with
increased sensitivity to the velocity `corners' may be found. To this end, we
tested a higher-order signal improvement scheme using principal component
analysis (PCA). In practice, we computed the even harmonic orders for the
(aligned) velocity field in annuli for even harmonic orders up to $m_{\rm
max}=10$, following
\begin{equation}
    v_{m\perp}=\frac{1}{N}\sum_{i=0}^N v_{\perp,i}\cos\left(m\phi_i\right),
    \label{eq:mvelocity}
\end{equation}
and constructed the matrix
\begin{equation}
\mathbfss{X}=\left[v_{4\perp},v_{6\perp},v_{8\perp},v_{10\perp}\right]^\intercal,
\end{equation} where each $v_{m\perp}$ is a list of values in each radius bin.
We then decomposed the matrix $X$ using singular value decomposition,
\begin{equation}
    \mathbfss{X} = \mathbfss{U}{\bf \Sigma}\mathbfss{V}^\intercal.
\end{equation}

The columns of $\mathbfss{U}$ are the principal components, the first of which
encodes the linear combination that makes up the strongest signal across the
whole galaxy face\footnote{By limiting the radius at which we evaluate $v_{m\perp}$, one could in principle isolate the strongest signal to be from the bar, but here we do not restrict the radius of evaluation.}. The PCA reveals that the
signal is indeed dominated by the $m=4$ signal, but higher order moments can
contribute, particularly when the geometry of the bar is long and narrow. In the
case of the primary model studied in this paper, $v_{8\perp}$ contributes to the
maximal signal at approximately the 20 per cent level during the assembly and
growth phases -- when the bar is particularly long and narrow. This may be
thought of as a signal `boost'. We also checked for signal in the sine terms
with the same processes (i.e. replacing cosine with sine in
eq.~\ref{eq:mvelocity}), as well as odd $m$ terms, and found no appreciable
contribution. One could also test a cumulative power ratio strategy, where all
non $m=2$ power is combined as an indicator of $x_1$ orbits.

Overall, we find only marginal improvement against the naive
$v_{4\perp}/v_{2\perp}$ ratio, but for cases where the bar is particularly
narrow, or the signal is weak, such a PCA process may prove useful. We expect
such a technique may be more useful when measuring signals in real galaxies,
where the inclination and observational uncertainty may obscure the signal.

\section{Detailed Ellipse Fit Demonstration} \label{sec:ellipsefitdemo}

\begin{figure*} \centering \includegraphics[width=6.5in]{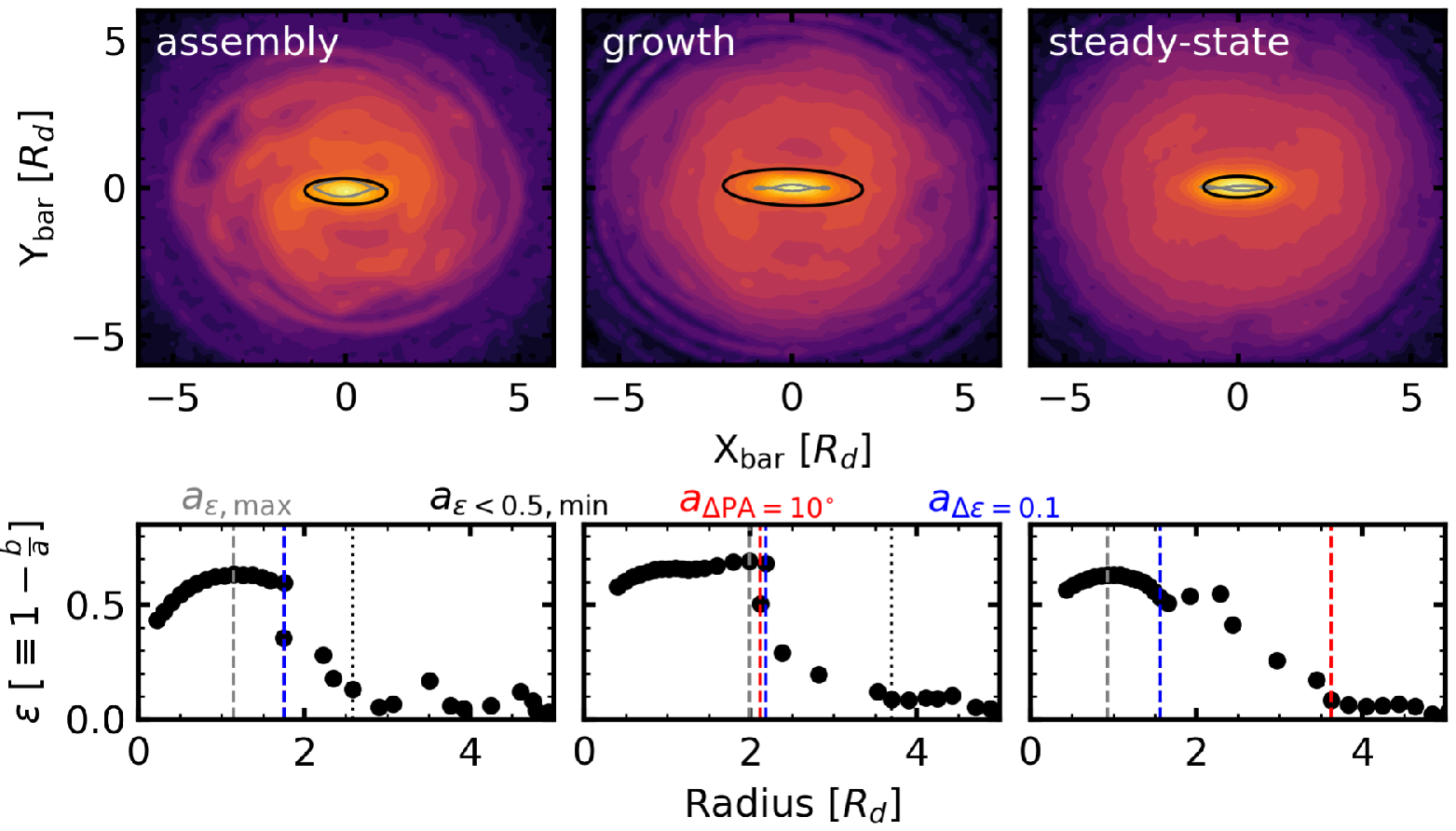}

\caption{\label{fig:literatureellipses} Upper row: three face-on surface
density snapshots extracted from the simulation, at different phases of bar
evolution. Lower row: eccentricity vs major axis for ellipses drawn to match
isophotes in the surface density snapshots. We also demonstrate different
ellipse-fitting methods and the results for bar length (grey, black, red, and
blue vertical lines, using the methods indicated above the panels).} \end{figure*}

Figure~\ref{fig:literatureellipses} shows ellipse-fit diagnostics for the
primary simulation considered in this paper. This figure emphasises the
ambiguity when performing ellipse fits, rather than the conceptually well-posed
$x_1$ velocity signal. In each face-on image, we also show the maximal
$x_1$ orbit identified by the $k$-means classifier, relative to the maximum-
ellipticity ellipse. In some cases, the maximum-ellipticity ellipse is a good
match, but this appears to be purely coincidental, as demonstrated by the poor
match during the growth phase.

\section{A sample of isolated galaxy simulations}
\label{sec:samplegalaxies}

\begin{figure*} \centering \includegraphics[width=6.5in]
{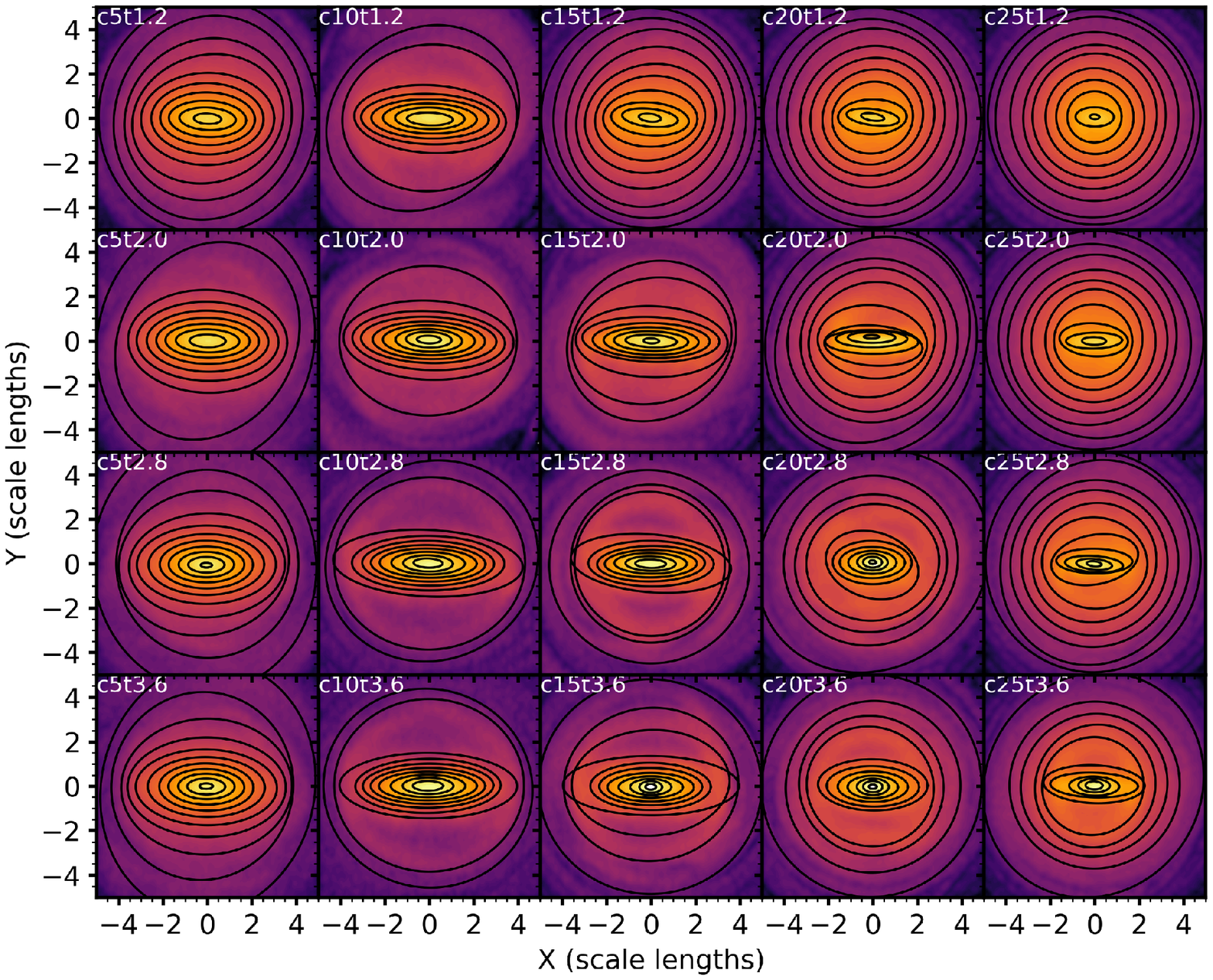}
\caption{\label{fig:barsnapshots} Face-on snapshots of five different galaxy
models (columns) at four different times (rows). The models are described in the
text and are designed to show a range of bar and disc morphologies in
isolated galaxies. On each panel, we overlay a handful of indicative ellipse
fits to the isophotes, as black curves.} \end{figure*}

\begin{figure} \centering \includegraphics[width=3.0in]{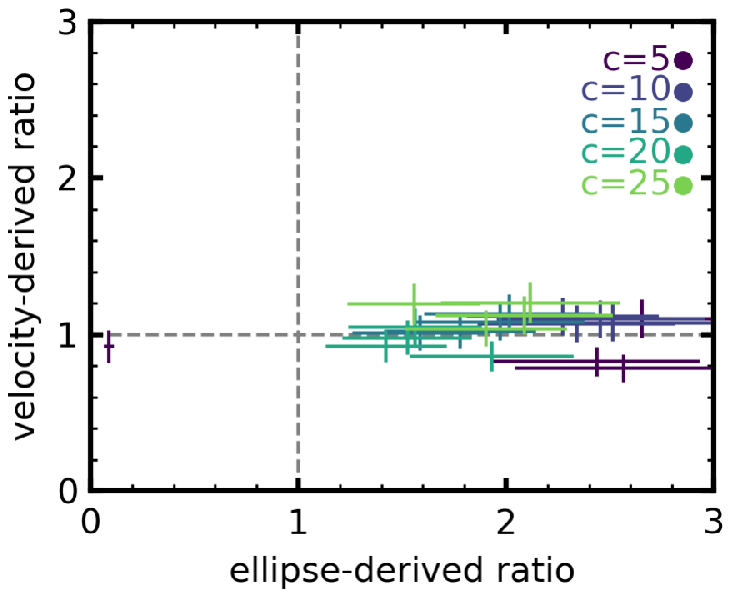}
\caption{\label{fig:sampleratio} The ratio of the maximum-ellipticity semimajor
axis to the determined maximal $x_1$ length (`ellipse-derived ratio') versus the
ratio of the $x_1$ velocity derived length to the maximal $x_1$ length
(`velocity-derived ratio'). The velocity-derived ratio consistently returns a
value near the maximal $x_1$ length. Different colour points correspond to
different models (darkest are $c5$, lightest are $c25$, marked in the legend).
Each point carries an uncertainty estimate derived from repeated observations
while changing parameters; the uncertainty in the ellipse-derived ratio is
approximately 20 per cent and the uncertainty in the velocity-derived ratio is
approximately 10 per cent.} \end{figure}

To probe a wider range of bar models, and test the robustness of our technique
to measure the dynamical length, we generate a new sample of barred galaxies by
varying the concentration in a NFW \citep{navarro97} dark matter halo. The dark
matter density affects the formation, structure, and evolution of bars in
galaxies \citep[see, e.g.][]{debattista00}. We effectively increase and decrease
the dark matter density in the disc region by changing the concentration of the
dark matter halo compared to the original model of
\citet{Petersen..commensurabilities..2021}, which had $c=15$, using $c=[5,10,15,20,25]$
to span a very large range of NFW halo models. The models are denoted by their
concentration, e.g. $c5$ for the $c=5$ model.

The models follow the initial condition generation procedure described in
\citetalias{Petersen..commensurabilities..2021}, with the only change being the
scale height: in the newly-generated sample we choose a ratio of $z_d/R_d =
0.15$, in contrast to the primary model presented in this paper, which has
$z_d/R_d = 0.1$. The increase in scale height of the model makes the disc more
in line with estimates for the structure of the Milky Way thin disc
\citep[$h_{\rm scale}=0.0015$ in virial units, or 450 pc when scaled to the
Milky Way, cf.][]{Petersen..commensurabilities..2021}. All new models are run
for $T_{\rm vir}=5$, using the same settings as in
\citetalias{Petersen..commensurabilities..2021} for the cusp model. Each model
forms a bar that then evolves. The bars span a relatively broad set of
morphologies: from shorter more compact bars to highly elongated bars. We track
the length of the $x_1$ orbit family in the live simulation in the same  manner
as described in the main text.

Figure~\ref{fig:barsnapshots} shows each galaxy model ($c5$, $c10$, $c15$, $c20$,
$c25$, in columns) in snapshots at four times ($T=1.2$, $T=2.0$, $T=2.8$,
$T=3.6$, in rows). The snapshots demonstrate the breadth of the bar morphologies
that we test. For each of these twenty snapshots, we perform an ellipse fit to
the face-on image. A selection of resultant ellipses are shown overlaid on each
image. From a full set of ellipse fits, we determine the semi-major axis of the
maximum ellipticity ellipse, and compute the ratio of this semi-major axis to
the maximal \(x_1\) length as the `ellipse-derived ratio'. We plot these values
as the $x$-axis of Figure~\ref{fig:sampleratio}, and show they consistently
overestimate the bar length (excepting one failed ellipse-determination),
with a mean overestimate of 2x. As an estimate of the uncertainty, we take the
semimajor axis of the ellipse fit to a log10 isophotal value that is 10 per cent
larger. This results in an approximately 20 per cent uncertainty estimate on the
ellipse-derived length. The uncertainty in the maximal \(x_1\) length is much
smaller than the ellipse semi-major axis uncertainty and, therefore, we only
indicate the ellipse uncertainty in Figure~\ref{fig:sampleratio}.

We also perform the Fourier-velocity bar length measurement for each sample
snapshot, and compute the ratio of the derived length to the $x_1$ length. This
is the $y$-axis in Figure~\ref{fig:sampleratio}. Here, we see that the
measurements are clustered around unity, indicating that the velocity method
recovers the $x_1$ length in this sample of snapshots. As an estimate of the
uncertainty, we follow the same procedure as in the main text and compute the
minimum for a variety of bin sizes and numbers. We then take the points and fit
a negative Gaussian, using the standard deviation of the fit as the uncertainty
on the velocity-derived length. This typically yields 10 per cent uncertainties.

Overall, we find that the velocity-derived metric consistently returns a bar
length that is comparable to the dynamical length. We acknowledge that the bar
morphology in these models is not necessarily a representative sample of the
observed morphologies found in the real Universe. Future work could additionally
test our technique on cosmological simulations, where bars may be directly
caused by interactions, resulting in a broader set of morphologies. However, our
snapshots do demonstrate that the velocity method works for a wide range of
isolated barred galaxy models, with essentially zero failures, even in the case
of very weak bars (e.g. $c25t1.2$).

\end{document}